\documentclass[reprint,aps,amsmath,amssymb,amsfonts,showkeys]{revtex4-2}
\usepackage[dvipdfmx]{graphicx}
\usepackage{dcolumn}
\usepackage{bm}
\usepackage{times}
\usepackage{multirow}
\usepackage{color}

\usepackage{ulem}
\newcommand{\rc}[1]{\textcolor{black}{#1}}

\begin{document}

\title{Transverse thermoelectric conversion in the mixed-dimensional semimetal WSi$_2$}

\author{Shoya~Ohsumi$^1$}
\email{6224508@ed.tus.ac.jp}
\author{Yoshiki~J.~Sato$^{1,2}$}
\author{Ryuji~Okazaki$^1$}

\affiliation{$^1$Department of Physics and Astronomy, Tokyo University of Science, Noda 278-8510, Japan}
\affiliation{$^2$Graduate School of Science and Engineering, Saitama University, Saitama 338-8570, Japan}

\begin{abstract}
Materials with axis-dependent conduction polarity are known as $p\times n$-type or goniopolar conductors that can be used for transverse thermoelectric devices, allowing the longitudinal thermal current to be converted into the transverse electrical current.
Here, we performed experimental and computational studies on the transport properties of WSi$_2$ single crystals, in which such axis-dependent conduction polarity of the thermopower and the Hall coefficient have recently been reported, and demonstrated the transverse thermoelectric effect by applying the temperature gradient in a 45$^{\circ}$ rotated direction from the crystallographic axis. 
We have observed strongly sample-dependent transport properties, which have also been observed in previous studies, and together with first-principles calculations we show that such sample-dependent transport properties originate from the band-dependent scattering rates of carriers.
The calculated band-resolved Peltier conductivity shows that the mixed-dimensional electronic structure consisting of a quasi-one-dimensional electron Fermi surface and a quasi-two-dimensional hole surface is a key property for the axis-dependent conduction polarity.
The directly obtained transverse thermoelectric figure of merit is comparable to that of the anomalous Nernst materials, implying that the present material is a potential candidate for transverse thermoelectric conversion.
\end{abstract}

\maketitle

\section{introduction}

Unlike conventional thermoelectric devices, which convert thermal current into electrical current and vice versa in the parallel direction, the transverse thermoelectric (TTE) device enables a perpendicular-induced electrical current against the temperature gradient, achieving efficient thermoelectric conversion without significant loss at many electrical contacts \cite{Jandl1994,goldsmid2011application,zhou2013driving,tang2015p,Grayson18,Wang2020,Uchida2022}.
The efficiency of the TTE device is determined by the transverse dimensionless figure of merit $z_{xy}T$. When the heat current flows along the $x$ axis and the electromotive force is induced along the perpendicular $y$ axis, $z_{xy}T$ is defined using the material parameters as $z_{xy}T=S_{xy}^{2}T/\rho_{yy}\kappa_{xx}$, where
$S_{xy}$ is the transverse thermopower, which is the off-diagonal term of the thermopower tensor $\hat{S}$, $\rho_{yy}$ is the electrical resistivity along the $y$ axis, $\kappa_{xx}$ is the thermal conductivity along the $x$ axis, and $T$ is the temperature \cite{goldsmid2011application,zhou2013driving}.
The well-known mechanism of the transverse thermoelectric effect is the Nernst effect in a magnetic field, and various topological magnets are focused as potential TTE materials \cite{sakai2018giant,guin2019anomalous,sakai2020iron,asaba2021colossal,Uchida2021APL,pan2022giant,Roychowdhury2022,Pan2022WTe2}.
In addition, the transverse thermopower appears even in a zero magnetic field in anisotropic materials:
when the $x$-axis direction is rotated by $\theta$ from the crystallographic $i$ axis to the $j$ axis, $S_{xy}$ is given as $\left(S_{ii}-S_{jj}\right)\sin\theta\cos\theta$, where $S_{ii}$ and $S_{jj}$ are the thermopower along the crystallographic $i$ and $j$ axes, respectively. Thus, $p\times n$-type conductors \cite{zhou2013driving}, which have the axis-dependent conduction polarity (ADCP), and goniopolar materials \cite{He2019}, in which a unique band curvature contributes to ADCP, can enhance $S_{xy}$ and then these materials are great candidates for TTE materials \cite{kawasugi2016simultaneous,Scudder2021,Ochs2021,nakamura2021axis,Scudder2022,Koster2023,Nelson2023,Goto2024}.
In addition, mixed-dimensional materials, in which both one-dimensional and two-dimensional Fermi surfaces coexist with opposite polarities, are anticipated to induce the ADCP \cite{manako2024}.

To demonstrate the large TTE conversion in the materials with ADCP, here we focus on the semimetal WSi$_{2}$, which crystallizes in the tetragonal crystal structure of the space group $I4/mmm$ [Fig. 1(a)]. This material exhibits extremely large magnetoresistance, which is interpreted as the result of the nearly perfect compensation of electrons and holes with high mobility \cite{mondal2020extremely,pavlosiuk2022giant}. Importantly, this material exhibits ADCP in both the Hall effect and the thermopower \cite{mondal2020extremely,Koster2023}. The Fermi surfaces have been well studied by means of quantum oscillation measurements and the first-principles calculations, suggesting that the hole is the dominant carrier in the in-plane conduction and the electron is dominant for the out-of-plane conduction along the $c$-axis direction \cite{mondal2020extremely,pavlosiuk2022giant}. However, the thermopower measurements have been performed above 80~K and therefore it remains to be determined whether the low-temperature ADCP originates from anisotropic Fermi surfaces, because the ADCP often depends on temperature as seen in several materials \cite{Rowe1970,Cohn}. Moreover, the origin of the observed strong sample-dependent transport properties is not clear. Furthermore, the anticipated TTE effect from this ADCP has not been directly detected in experiments.

In this study, we measured the thermopower, electrical resistivity, and thermal conductivity of WSi$_{2}$ single crystals along the $a$ and $c$ axes at low temperatures. We find that this material exhibits the ADCP of the thermopower at low temperatures and shows the sample-dependent transport properties along the $c$ axis. First-principles calculations were also performed, and we show that the ADCP originates from the mixed-dimensionality of the Fermi surfaces, which is characterized by the coexistence of a quasi-one-dimensional electron-like and a quasi-two-dimensional hole-like Fermi surfaces. Based on the calculated band-dependent transport properties, the observed sample-dependent transport properties can be explained by the band-dependent relaxation time.
We also directly demonstrated TTE generation by applying a temperature gradient along the 45$^{\circ}$ rotated direction from the $a$ axis to the $c$ axis. The electromotive force perpendicular to the temperature gradient is clearly detected, and the results are consistent with the estimated values from the thermopower along the crystallographic axes. These results are a good trigger for accelerating the exploration of the ADCP materials and realization of the TTE conversion using the ADCP.

\begin{figure}[t]
\begin{center}
\includegraphics[width=8cm]{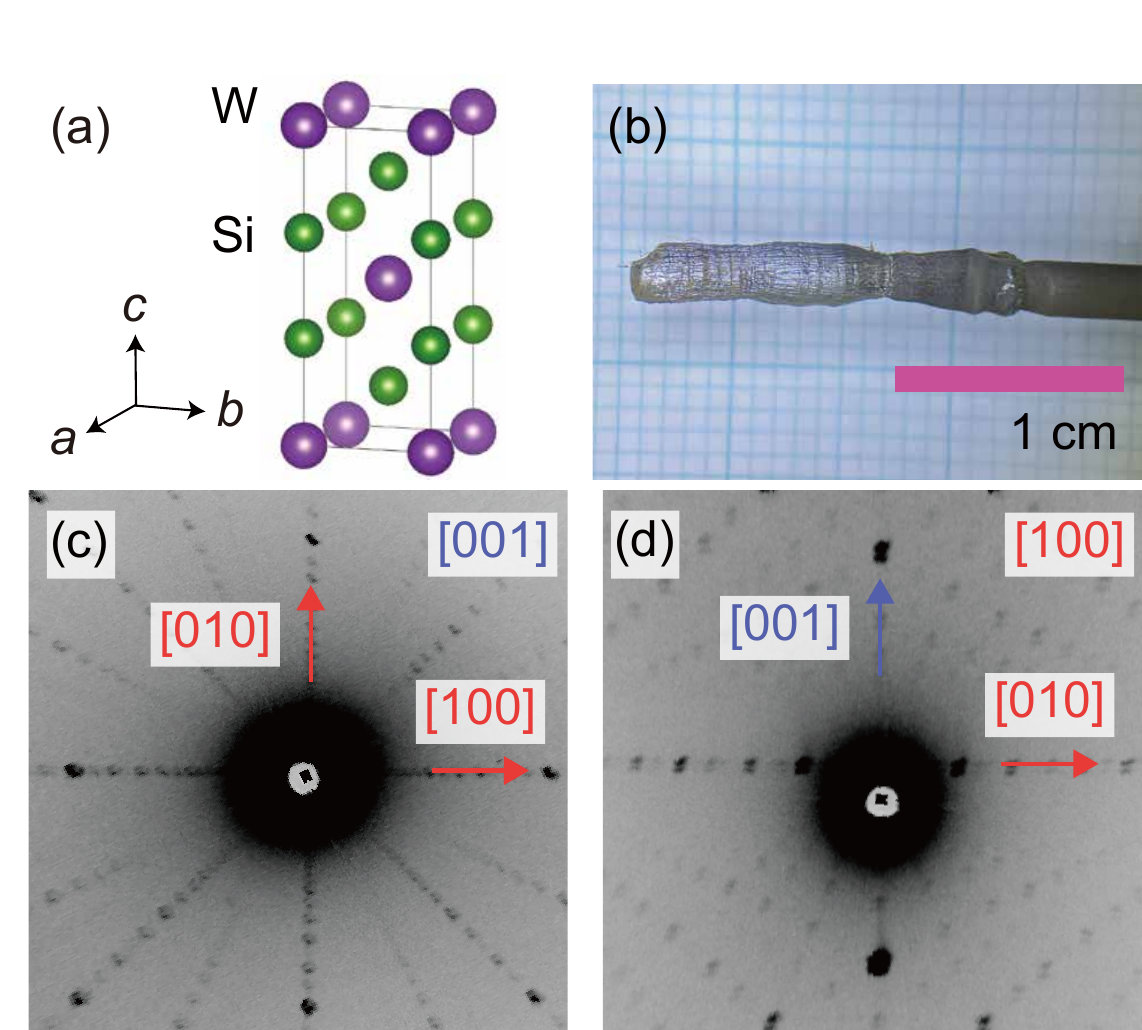}
\caption{
(a) Crystal structure of WSi$_{2}$ drawn by VESTA \cite{Momma2011}.
(b) Single-crystalline WSi$_{2}$ sample grown using the Czochralski method. 
(c,d) Laue backscattering X-ray diffraction pattern of WSi$_{2}$ single crystal with X-ray beam along [001] direction (c) and [100] direction (d).
}
\end{center}
\end{figure}

\begin{figure}[t]
\begin{center}
\includegraphics[width=8cm]{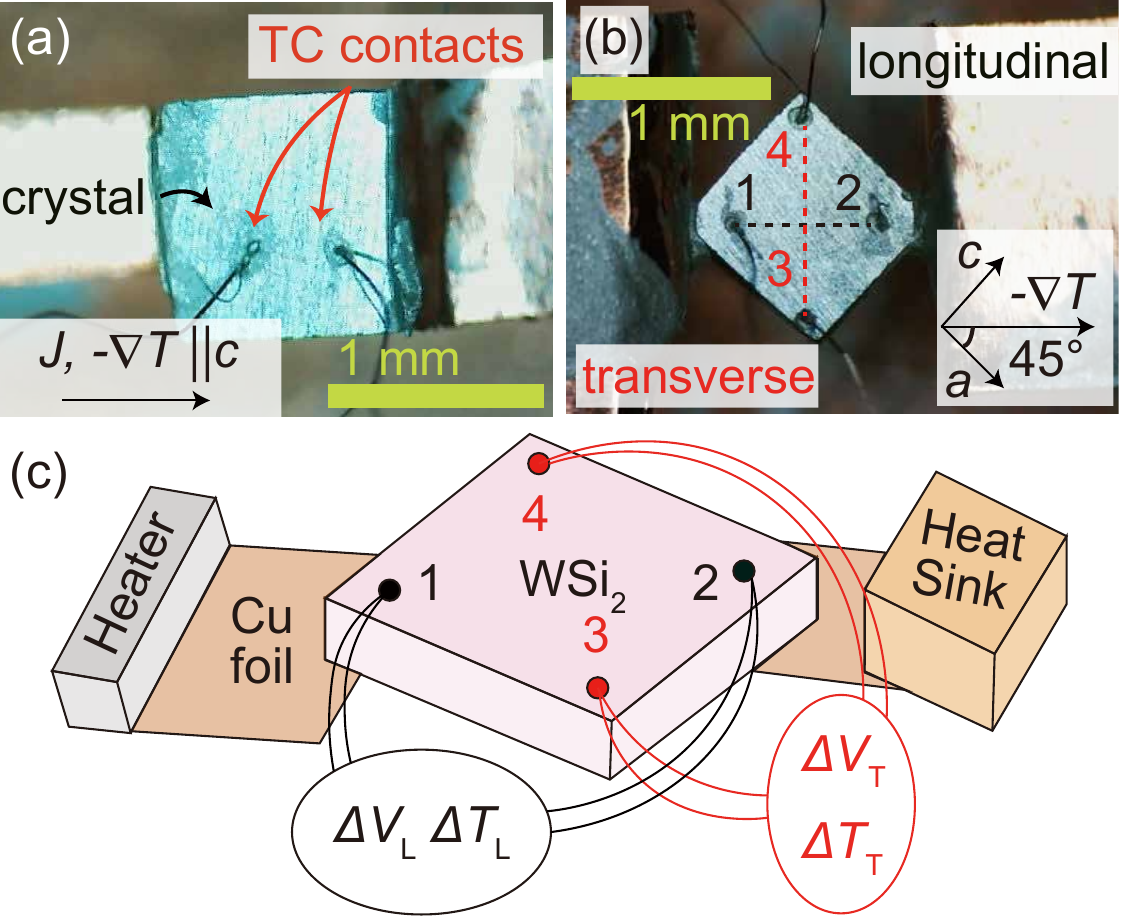}
\caption{
(a) Measurement setup for the transport properties along the crystallographic direction.
The electrical current density $\bm{J}$ and the temperature gradient $\bm{\nabla}T$ are applied along the $c$-axis direction. The temperature different is measured using two thermocouples (TCs).
(b) Measurement setup for the TTE effect.
The temperature gradient $\bm{\nabla}T$ in a 45$^{\circ}$ rotated direction from the crystallographic axis. 
The voltage and temperature difference along the longitudinal (black dashed line) and transverse (red dashed line) directions are measured using four TCs.
(c) Schematic representation of measurement setup for the TTE effect shown in (b). In this geometry, the longitudinal voltage difference $\Delta V_{\mathrm{L}}$ and temperature difference $\Delta T_{\mathrm{L}}$ are measured using the TCs connected at 1 and 2, and the transverse voltage difference $\Delta V_{\mathrm{T}}$ and temperature difference $\Delta T_{\mathrm{T}}$ are measured using the TCs connected at 3 and 4.
}
\end{center}
\end{figure}

\section{Method}
\subsection{Crystal growth}
Polycrystalline sample was first synthesized. Starting materials with the ratio of $\mathrm{W:Si}=1:2.1$ (5$\%$ excess) were melted in a tetra-arc furnace under Ar atmosphere. The ingot has been melted six times by flipping upside down before each melt to improve the homogeneity. Then, the WSi$_2$ single crystal was grown using the Czochralski method in a tetra-arc furnace under Ar atmosphere \cite{mondal2020extremely}. The single crystal was pulled at a rate of 15 mm/h for about 3 h. The photograph of the grown crystal is shown in Fig. 1(b). 
To check the chemical compositions, we performed electron probe micro analyzer (EPMA) measurements. As shown in Fig. S5 in the Supplemental Material \cite{supplemental}, the Si content is slightly low from the stoichiometric value, probably due to the Si deficiency, while we added the excess Si before the melting process.
In order to orient the measured single crystals along the crystallographic axis and check their quality, the backscattering Laue method was carried out as shown in Figs. 1(c) and 1(d). We measured 4 samples $\#1-\#4$. The samples $\#1-\#3$ were cut from different parts of the same single-crystalline batch, while the samples $\#4$ were cut from a different batch. We measured the transport properties of the sample $\#3$ before and after annealing. The data of the samples $\#3$ and $\#4$ are shown in the Supplemental Material \cite{supplemental}.

\subsection{Transport measurement}
Electrical resistivity, thermopower and thermal conductivity were measured simultaneously using a hand-made probe from 10 to 300 K \cite{Takahashi2016}.
Note that some data at low temperatures are not available due to the significant thermal conductance of crystals.
Electrical resistivity was measured using a standard four-probe method with a Keithley 2182A nanovoltmeter. The electrical current of $I=3\,\mathrm{mA}$ was provided by a Keithley 6221 current source. Thermal conductivity and thermopower along the crystallographic axes were measured using a steady-state method with manganin-constantan differential thermocouples as shown in Fig. 2(a). A temperature gradient of about 0.5 K/mm was applied using a resistive heater, the resistance of which was also measured using a four-probe method to calculate the applied heater power. The thermoelectric voltage of the wire leads was subtracted.
To measure the TTE effect, we applied the temperature gradient in a 45$^{\circ}$ rotated direction from the $a$ and $c$ axes as shown in Fig. 2(b). Also, the schematic representation is shown in Fig. 2(c).
The transverse and longitudinal voltage and temperature differences were measured simultaneously using four manganin-constantan thermocouples. The longitudinal voltage and temperature differences were measured by the two thermocouples connected at points 1 and 2, while the transverse voltage and temperature differences were measured by the two thermocouples connected at points 3 and 4.

The error of the resistivity was determined to be 10 $\%$, which was calculated based on the ratio of the diameter of the silver epoxy used to affix thermocouples to the sample to the length between the thermocouple contacts. The error of the thermal conductivity was estimated taking the geometrical error and the heat loss into account.  The geometrical error was determined to be 10 $\%$, which is the same as the electrical resistivity, and the heat loss was estimated using the thermal resistance of the sample and lead wires.

\subsection{First-principles calculations}
We performed first-principles calculations based on density functional theory using QUANTUM ESPRESSSO \cite{giannozzi2009quantum,giannozzi2017advanced}. We used the Perdew-Burke-Ernzerhof generalized-gradient-approximation (PBE-GGA) exchange-correlation functional. The cutoff energies for plane waves and charge densities were set to 90 and 720 Ry, respectively, and the dense $k$-point mesh was set to a uniform grid of $30\times30\times30$ to ensure the convergence. 
We used the on-site Coulomb energy $U=2.5$~eV and the exchange parameter $J=0.5$~eV for W ion \cite{Solovyev1994} and performed fully relativistic calculations with spin-orbit coupling (DFT+$U$+SOC). The present calculations are not spin-polarized.
We also calculated the transport coefficients based on the linearized Boltzmann equations under the constant-relaxation time approximation (Appendix A). 

\section{Results and discussion}

\begin{figure*}[t]
\begin{center}
\includegraphics[width=17cm]{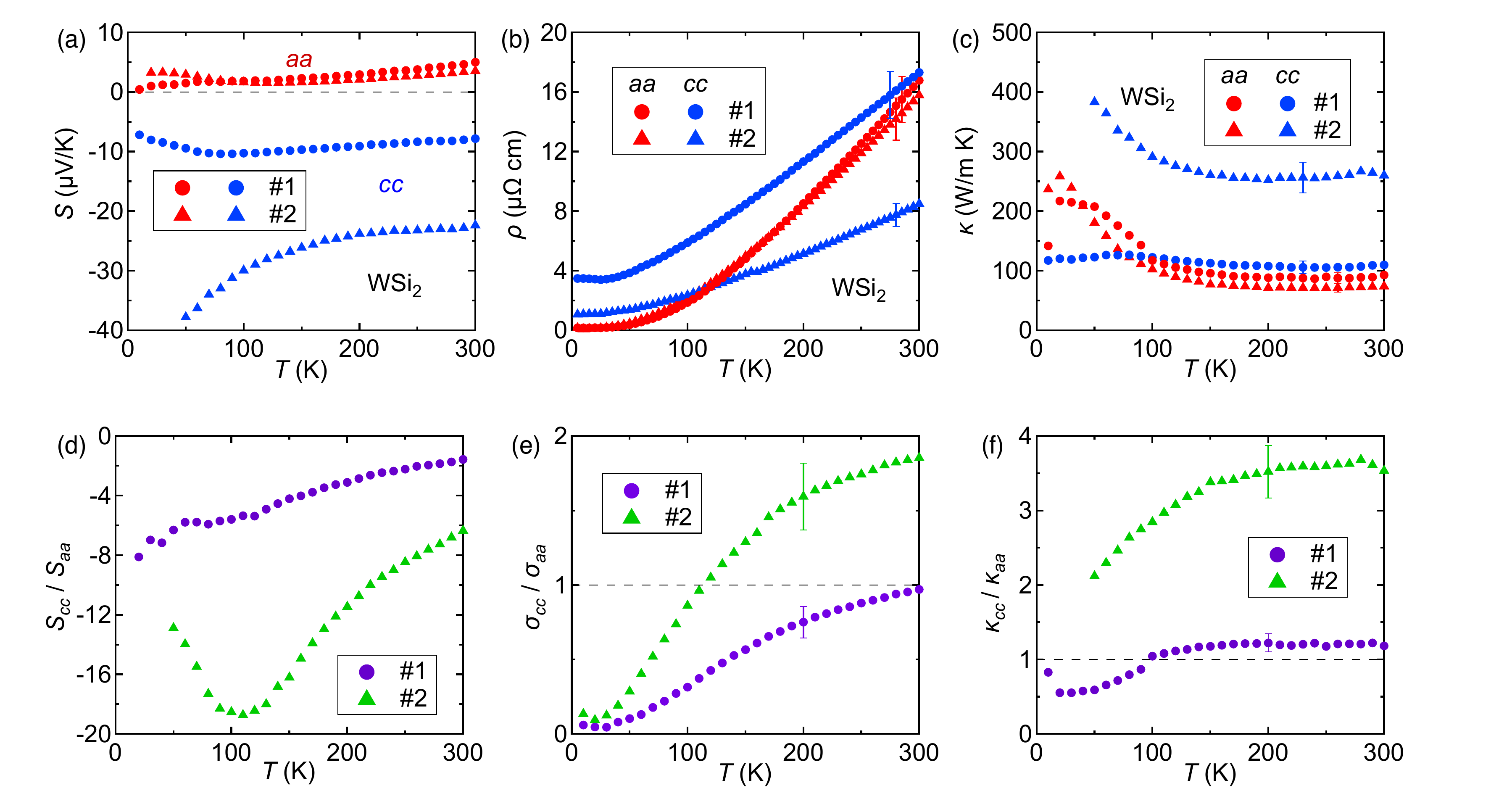}
\caption{
(a) Temperature dependence of the thermopower $S_{ii}$ of WSi$_2$ single crystals measured along the in-plane  ($i=a$) and out-of-plane  ($i=c$) directions.
(b,c) Temperature dependence of (b) electrical resistivity $\rho_{ii}$ and (c) thermal conductivity $\kappa_{ii}$. The error bars 
 of the resistivity and thermal conductivity represent the evaluated systematic error.
(d-f) Temperature dependence of the anisotropies in (d) thermopower, (e) electrical conductivity, and (f) thermal conductivity. The error bars represent the systematic error evaluated by the error propagation.
}
\end{center}
\end{figure*}

\subsection{Axis-dependent transport properties}

Figure 3 summarizes the anisotropic transport properties of WSi$_2$ single crystals. The results of the samples $\#3$ and $\#4$ are shown in Figs. S1 and S2 in the Supplemental Material \cite{supplemental}.
Figure 3(a) represents the temperature dependence of the thermopower $S_{ii}$ along the in-plane  ($i=a$) and out-of-plane  ($i=c$) directions for two measured single crystals, and ADCP with positive in-plane and negative out-of-plane thermopower is clearly resolved, as reported in the previous study \cite{Koster2023}.
The temperature dependence of $S_{ii}$ is nonmonotonic and different from the simple metallic behavior of $S\propto T$. $S_{cc}$ of both samples increase toward zero with increasing temperature above 100 K, indicating that the electron dominates the transport along the out-of-plane direction, while hole excitation occurs at high temperatures and contribution of electron and hole gradually compensate. The similar temperature dependence of thermopower was also measured in semimetallic materials \cite{fu2018large,liu2022weyl,scott2023doping,nakano2021giant}.

While the temperature dependence of the in-plane thermopower $S_{aa}$ exhibits a small sample dependence, the out-of-plane thermopower $S_{cc}$ shows a notable sample dependence.
Figures 3(b) and 3(c) show the electrical resistivity $\rho_{ii}$ and the thermal conductivity $\kappa_{ii}$ for both directions of the crystallographic axis, respectively. Resistivity shows conventional metallic behavior with electron-phonon scattering characterized by the low-temperature $T^5$ to high-temperature $T$ dependence.
Sample-dependent electrical resistivity and thermal conductivity were also observed, as is similar to the thermopower results. The out-of-plane components show significant sample dependence. We note that the non-monotonicity of thermopower cannot be explained by the electron-phonon coupling. As shown in Fig. S6 in the Supplemental Material \cite{supplemental}, the electron-phonon coupling is too weak to reproduce our experimental result, which supports the semimetallic origin of the thermopower. 
We also confirmed the Wiedemann-Franz law and obtained the Lorenz number $L=\kappa_{ii}/\sigma_{ii}T$, which is roughly consistent with the previous study \cite{Koster2023,koster2024giant}, and find that the contribution of the phonon to the thermal conductivity is dominant in the out-of-plane direction (see Fig. S7 in Supplemental Material \cite{supplemental}).

In Figs. 3(d-f), we evaluate the anisotropy of the measured transport properties. 
For the sample $\#1$, both the anisotropy value of the electrical conductivity $\sigma_{cc}/\sigma_{aa} (= \rho_{aa}/\rho_{cc})$ [Fig. 3(e)] and the thermal conductivity $\kappa_{cc}/\kappa_{aa}$ [Fig. 3(f)] are close to unity near room temperature. Note that $\sigma_{ii}$ and $\kappa_{ii}$ are electrical conductivity and thermal conductivity along the $i$ axis. On the other hand, these anisotropy values are larger than unity for the sample $\#2$, indicating that the observed sample-dependent resistivity and thermal conductivity originate from the sample-dependent anisotropy in the relaxation time in common, because $\sigma_{cc}/\sigma_{aa}$ and $\kappa_{cc}/\kappa_{aa}$ are proportional to the relaxation-time anisotropy in a classical picture.
Note that here we only consider the electronic contribution to the thermal conductivity in this qualitative discussion.
The origin of the sample-dependent relaxation time anisotropy will be discussed in section III.
C along with the anisotropy in the thermopower $S_{cc}/S_{aa}$ shown in Fig. 3(d).

\begin{figure}[t]
\begin{center}
\includegraphics[width=8.5cm]{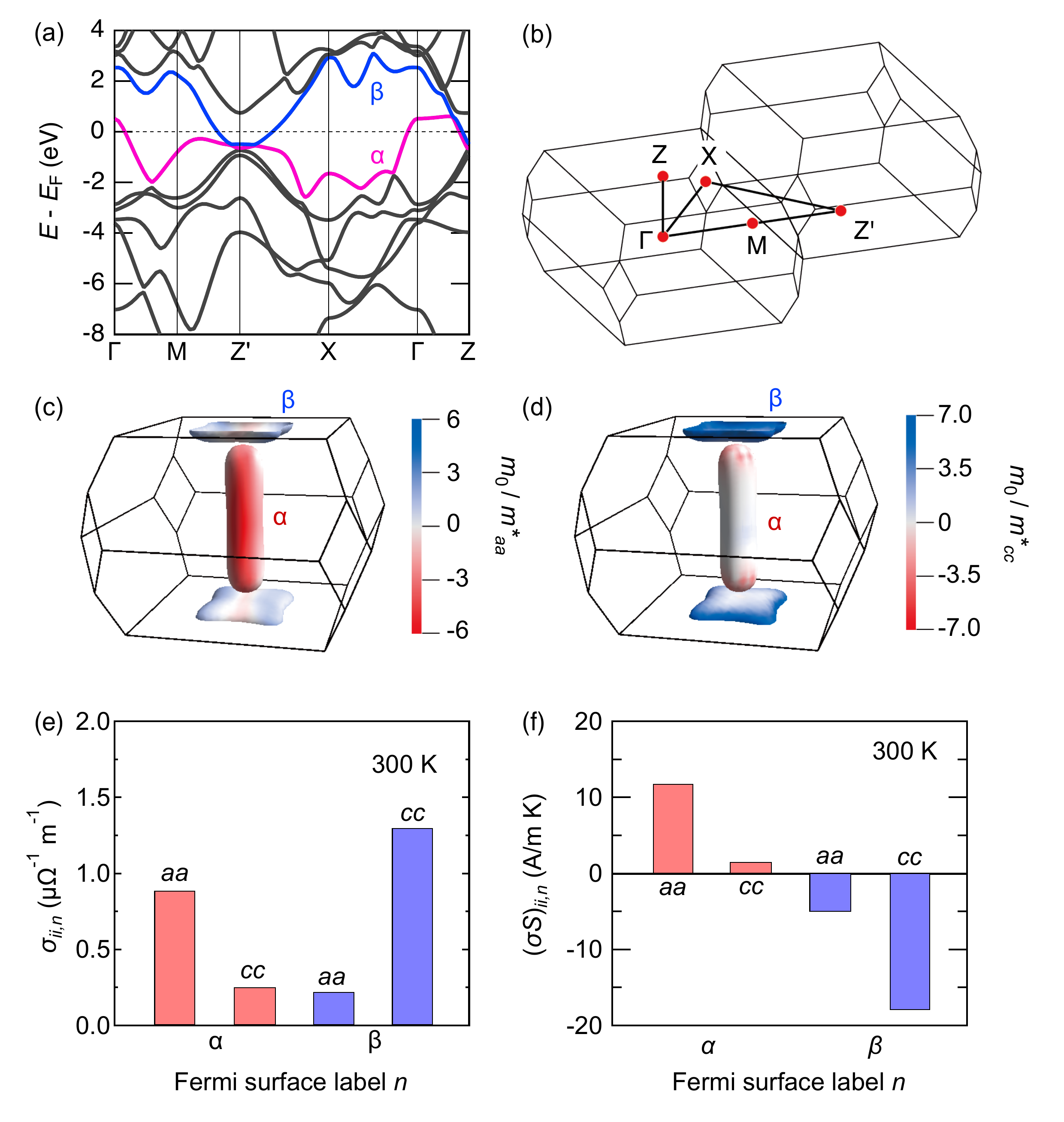}
\caption{
(a) Calculated band structure of WSi$_2$.
(b) High-symmetry points and the $k$ path for the band structure in (a).
$Z'$ is the $Z$ point in the adjacent Brillouin zone.
(c,d) The Fermi surfaces drawn by FermiSurfer \cite{Kawamura2019}. The color shows (c) in-plane  and (d) out-of-plane components of the inverse effective mass tensor.
The blue and red color indicates the electron- and hole-like dispersion at the Fermi level, respectively.
(e,f) Calculated band-resolved partial (e) electrical conductivity $\sigma_{ii,n}$ and (f) Peltier conductivity $(\sigma S)_{ii,n}$ ($i=a,c$ and $n=\alpha,\beta$). Data are calculated with a constant relaxation time of $10^{-14}$ s for $T=300$~K.
}
\end{center}
\end{figure}

\subsection{Calculated band structure and band-resolved transport properties}

\begin{figure*}[t]
\begin{center}
\includegraphics[width=16cm]{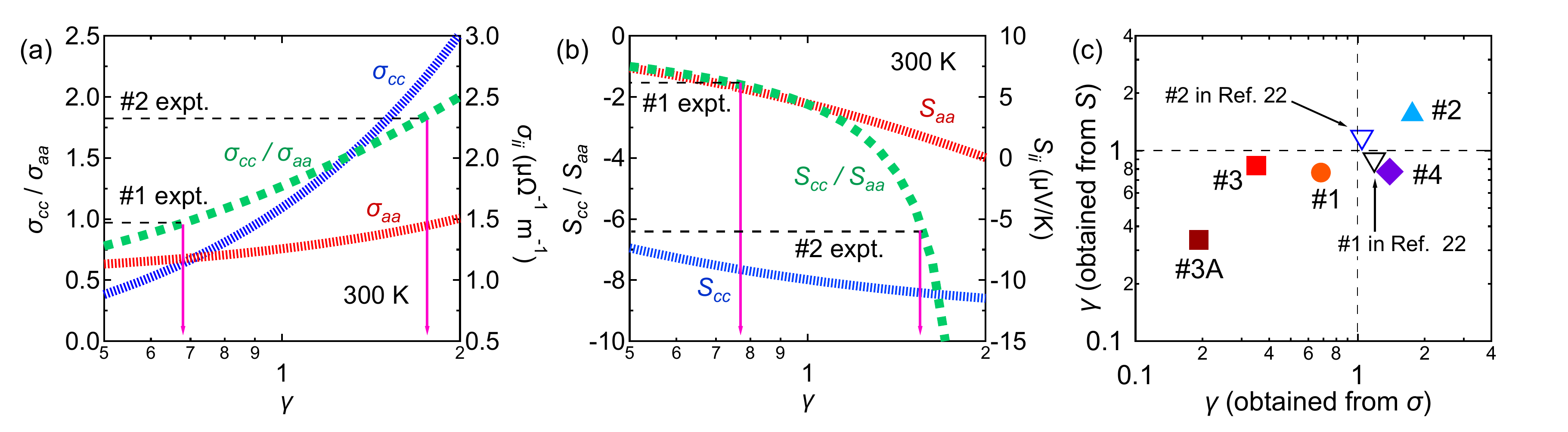}
\caption{
(a) Calculated total electrical conductivity $\sigma_{ii}$ ($i=a,c$, right axis) and its anisotropy $\sigma_{cc}/\sigma_{aa}$ as a function of the relaxation-time ratio $\gamma$.
Horizontal dashed lines represent the experimental values of $\sigma_{cc}/\sigma_{aa}$ for two measured samples and the corresponding values of $\gamma$ are indicated as vertical arrows.
(b) Calculated total thermopower $S_{ii}$ ($i=a,c$, right axis) and its anisotropy $S_{cc}/S_{aa}$ as a function of the relaxation-time ratio $\gamma$.
Horizontal dashed lines also represent the experimental values of $S_{cc}/S_{aa}$ and the corresponding values of $\gamma$ are indicated as vertical arrows.
(c) Relation between $\gamma$ obtained from the conductivity ratio and that obtained from the thermopower ratio. Samples $\#3$ and $\#3$A denote the experimental data of sample $\#3$ before and after annealing, respectively. We also analyzed the data in Ref. \cite{Koster2023}. 
}
\end{center}
\end{figure*}

To examine the origin of the sample-dependent transport properties, we discuss the calculated transport properties.
Figure 4(a) shows the band structure of WSi$_{2}$ depicted along the high-symmetry points shown in Fig. 4(b). As mentioned in previous studies \cite{mondal2020extremely,Koster2023}, the band structure of WSi$_{2}$ represents the coexistence of electron- and hole-like Fermi surfaces, which are shown in Figs. 4(c) and 4(d) with a color scale to show the in-plane component $m_0/m^*_{aa}$ and out-of-plane one $m_0/m^*_{cc}$ of the inverse effective mass tensor, respectively ($m_0$ being the electron rest mass).
The hole-like sheet named $\alpha$ sheet centered at the $\Gamma$ point has a  cylindrical shape extended in the $c$-axis direction which is regarded as a quasi-two-dimensional (q-2D) structure and dominates the in-plane conduction. 
In contrast, the electron-like sheet named $\beta$ sheet centered at the $Z$ point shows a large surface area with the normal vector along the $c$-axis direction which is then regarded as quasi-one-dimensional (q-1D), contributing to the out-of-plane transport property.
Hence, the coexistence of the q-2D hole-like $\alpha$ sheet and the q-1D electron-like $\beta$ sheet, which could be termed mixed-dimensionality, is responsible for the experimentally observed ADCP in WSi$_2$.
Such a mixed-dimensionality is a key property for the efficient TTE conversion \cite{manako2024}.

To quantitatively evaluate mixed-dimensional Fermi surfaces contribution to the anisotropic transport properties, we then calculate the band-resolved partial electrical and Peltier conductivity under a constant relaxation time approximation (Appendix A). 
Figures 4(e) and 4(f) show the partial electrical conductivity $\sigma_{ii,n}$, where $i=a,c$ is the crystallographic axis and $n=\alpha,\beta$ is the Fermi surface label, and Peltier conductivity $(\sigma S)_{ii,n}$, respectively, calculated with a constant relaxation time of $\tau_0=10^{-14}$ s for $T=300$~K. For the q-2D hole-like $\alpha$ sheet, the anisotropy of the electrical conductivity is estimated as $\sigma_{cc,\alpha}/\sigma_{aa,\alpha}\approx 0.2$ and the anisotropy of the Peltier conductivity yields $(\sigma S)_{cc,\alpha}/(\sigma S)_{aa,\alpha}\approx 0.3$, both of which are less than unity, quantitatively demonstrating that the sheet $\alpha$ dominates the in-plane conduction. 
On the other hand, the anisotropy of the q-1D electron-like $\beta$ sheet is evaluated as $\sigma_{cc,\beta}/\sigma_{aa,\beta}\approx 5.7$ and $(\sigma S)_{cc,\beta}/(\sigma S)_{aa,\beta}\approx 3.4$, dominating the out-of-plane conduction.

\subsection{Estimation of the sample-dependent anisotropy of relaxation time}
In this section, we elucidate the sample-dependent transport properties. Considering the semimetallic nature of this material, sample-dependent transport properties seem to originate from the chemical potential positions. However, the chemical potential position estimated from the anisotropic transport properties using first-principles calculations is inconsistent with the Hall measurement result (see Figs. S3 and S4 and Table S1 in the Supplemental Material \cite{supplemental}).
As mentioned above, the sample-dependent anisotropy in the relaxation time is essential to understand the strong sample-dependent transport properties in WSi$_2$. \rc{Indeed, evaluation of the relaxation time anisotropy is important to theoretically reproduce the thermopower in goniopolar materials and semimetal \cite{wang2020anisotropic,zhou2022anomalous}.}
Moreover, as demonstrated in the band-resolved calculations, the in-plane and out-of-plane transport properties originate mainly from the q-2D hole-like $\alpha$ and q-1D electron-like $\beta$ sheets, respectively, and therefore the relaxation time anisotropy can be regarded as a ratio of the band-dependent relaxation time $\tau_{\beta}/\tau_{\alpha}$, where $\tau_n$ ($n = \alpha,\beta$) is the relaxation time of $n$ sheet.

Now let us evaluate the band-dependent relaxation time ratio $\gamma=\tau_{\beta}/\tau_{\alpha}$ from the experimental data of thermopower and electrical conductivity as follows. \rc{Since WSi$_{2}$ is a typical semimetal and has two relatively simple Fermi surfaces, this material is a good example to evaluate the relation between the transport properties and the band-dependent relaxation time in multicarrier systems.}
Although $\gamma$ would be temperature-dependent, we discuss $\gamma$ at 300 K, where the diffusive part of thermopower is dominant \cite{markov2019thermoelectric}.

Here, we assume a constant relaxation time for the $\alpha$ sheet as $\tau_{\alpha} = \tau_0$  $(=10^{-14}$~s), and thus the relaxation time for the $\beta$ sheet is given as $\tau_{\beta} = \gamma\tau_{\alpha} = \gamma\tau_{0}$. Note that the absolute value of $\tau_0$ is not crucial in the present analysis, as we will discuss the conductivity ratio in the following. Thus, using $\gamma=\tau_{\beta}/\tau_{\alpha}$, the multiband description for the total electrical conductivity $\sigma_{ii}$ ($i=a,c$) reads
\begin{align}
\sigma_{ii}=\sigma_{ii,\alpha}+\gamma\sigma_{ii,\beta},
\label{cond}
\end{align}
where $\sigma_{ii,n}$ is the band-resolved electrical conductivity calculated with the constant relaxation time of $\tau_0=10^{-14}$~s as shown in Fig. 4(e).
Using the calculated values of $\sigma_{ii,n}$, we then simulate the total conductivity $\sigma_{ii}$ and its anisotropy $\sigma_{cc}/\sigma_{aa}$ as a function of $\gamma$ based on Eq. (\ref{cond}), which are plotted for the right and left axes of Fig. 5(a), respectively.
Subsequently, using the experimental values of $\sigma_{cc}/\sigma_{aa}$ at room temperature as depicted by the horizontal dashed lines in Fig. 5(a), we have obtained $\gamma = 0.68$ for sample $\#1$ and $\gamma = 1.76$ for sample $\#2$.
The higher value of $\gamma$ for sample $\#2$ than that for sample $\#1$ reflects that the electrical resistivity along the $c$-axis direction $\rho_{cc}$ of sample $\#2$ is lower than that of sample $\#1$ [Fig. 3(b)]; in sample $\#2$, longer relaxation time of carriers on the q-1D $\beta$ sheet may reduce the out-of-plane resistivity.

Similarly, the total thermopower in the multiband case is given as
\begin{align}
S_{ii}=\dfrac{\left(\sigma S\right)_{ii,\alpha}+\gamma\left(\sigma S\right)_{ii,\beta}}{\sigma_{ii,\alpha}+\gamma\sigma_{ii,\beta}},
\label{thermo}
\end{align} 
where $(\sigma S)_{ii,n}$ is the band-resolved Peltier conductivity calculated with the constant relaxation time of $\tau_0=10^{-14}$~s as shown in Fig. 4(f).
Then we simulate the total thermopower $S_{ii}$ and its anisotropy $S_{cc}/S_{aa}$ as a function of $\gamma$ as plotted in Fig. 5(b), and obtain $\gamma=0.77$ for sample $\#1$ and $\gamma=1.55$ for sample $\#2$, implying that the relaxation time of the $\beta$ sheet of sample $\#2$ is longer than that of sample $\#1$ similar to the results of the conductivity analysis.
Such a long relaxation time of carriers on the q-1D electron-like $\beta$ sheet negatively enhances the out-of-plane thermopower of sample $\#2$ [Figs. 3(a) and 3(d)].

In Fig. 5(c), we plot the relation between $\gamma$ obtained from the conductivity ratio and one from the thermopower ratio for samples $\#1$-$\#4$.
In addition, we estimate $\gamma$ of samples in the previous study \cite{Koster2023} and the results are also plotted in Fig. 5(c). A positive correlation between these $\gamma$ values is clearly observed, indicating that the ratio of the band-dependent relaxation time $\gamma = \tau_{\beta}/\tau_{\alpha}$ is indeed sample-dependent and affects the transport properties of WSi$_2$.
Note that the anisotropy of thermal conductivity is not analyzed because of the difficulty of separating the electron and phonon contributions.

\begin{figure}[t]
\begin{center}
\includegraphics[width=8cm]{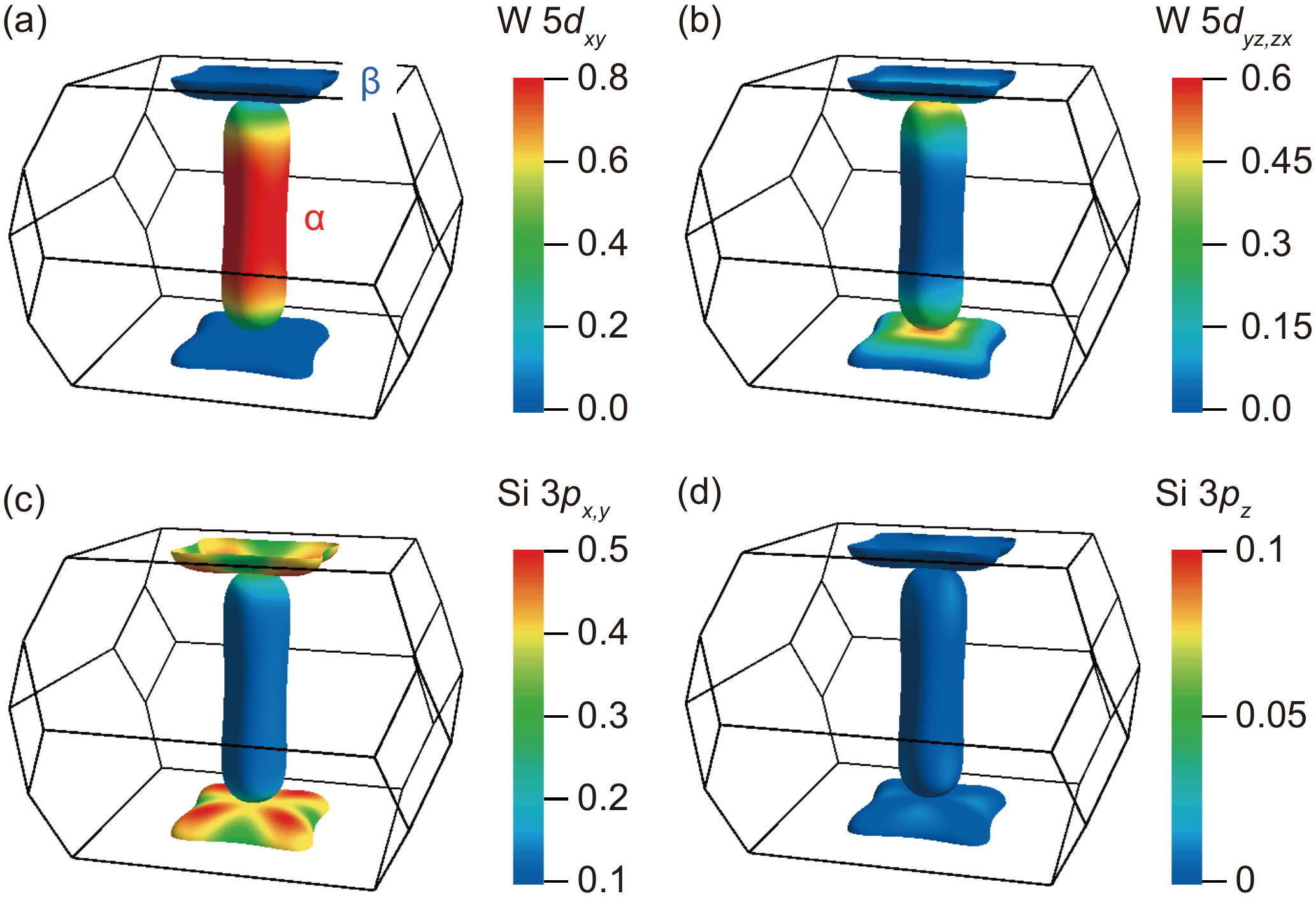}
\caption{
Orbital weights on the Fermi surfaces for (a) W 5$d_{xy}$, (b) W 3$d_{yz,zx}$, (c) Si 3$p_{x,y}$ and (d) Si 3$p_{z}$ orbitals. The color scale indicates the sum of the projection of the $m$ orbital $\sum_{m}\left|\left<\phi_{m}|\psi\right>\right|^2$ \cite{Kawamura2019}.}
\end{center}
\end{figure}

The origin of such a band-dependent scattering time is now discussed in terms of atomic orbitals in WSi$_2$.
In Figs. 6(a) and 6(b), we show the calculated orbital weights on the Fermi surfaces for W $5d_{xy}$ and $5d_{yz/zx}$ orbitals, respectively.
The hole-like $\alpha$ sheet for the in-plane conduction mainly originates from W $5d_{xy}$ orbital and the electron-like $\beta$ sheet for the out-of-plane conduction stems from W $5d_{xz/yz}$ orbitals, which has also been claimed in the band-structure form with a specific $k$ path \cite{Koster2023}. 
In addition, here we find that the orbital weights of Si $3p$ orbitals are band-dependent [Figs. 6(c) and 6(d)]; the Si $3p_{x,y}$ orbitals contribute significantly to the $\beta$ sheet.
Therefore, we speculate that defects and/or crystalline imperfections of the Si sites may strongly affect the relaxation time of carriers on the $\beta$ sheet, resulting in the observed sample-dependent out-of-plane transport properties.

\subsection{Demonstration of TTE conversion}

\begin{figure}[t]
\begin{center}
\includegraphics[width=8.9cm]{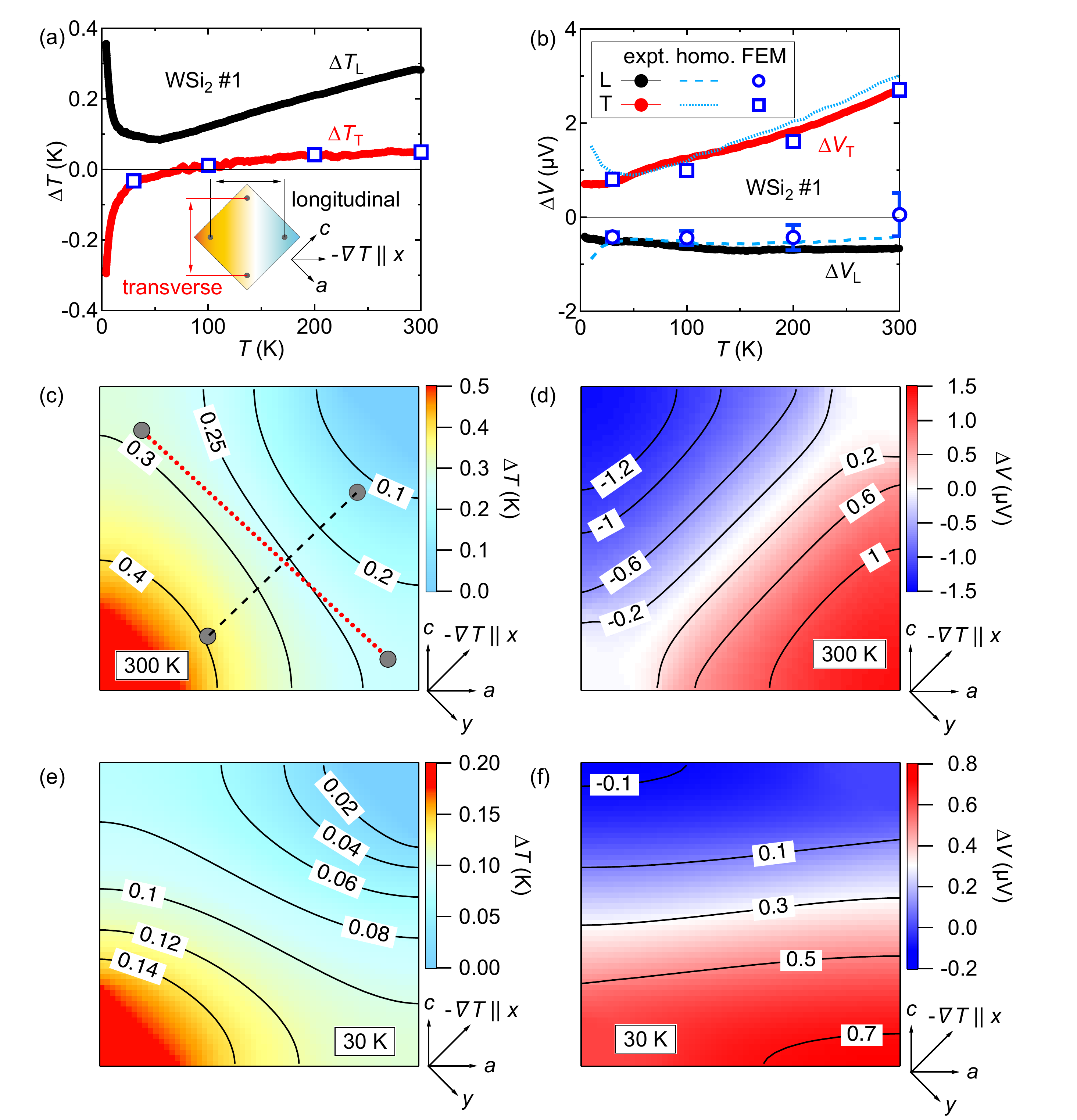}
\caption{Verification of the TTE conversion.
(a) Temperature dependence of the temperature difference along the longitudinal direction $\Delta T_{\mathrm{L}}$ and the transverse direction $\Delta T_{\mathrm{T}}$. The temperature differences were measured using the thermocouples connected at 1 and 2, and 3 and 4 shown in Fig. 2(b), respectively.
The open squares are calculated $\Delta T_{\mathrm{T}}$ using the finite element method (FEM).
(b) Temperature dependence of the potential difference along the longitudinal direction $\Delta V_{\mathrm{L}}=V_{12}$ and the transverse direction $\Delta V_{\mathrm{T}}=V_{34}$. 
The dashed and dotted curves show the estimated values using the Eqs. (3) and (4), respectively, calculated under the assumption of the homogeneous temperature gradient.
The calculated values for $\Delta V_{\mathrm{L}}$ and $\Delta V_{\mathrm{T}}$ using the FEM are also plotted with open circles and squares, respectively. 
(c,d) The temperature and the potential distribution at 300 K calculated using FEM. We consider a square-shaped sample as shown in Fig. 2(b) and the circles show the positions of the thermocouple contacts. The dashed and dotted lines denote the longitudinal and transverse directions, respectively.  
(e,f) Temperature and potential distribution at 30 K calculated using FEM.}
\end{center}
\end{figure}

Finally, we examine the TTE generation in WSi$_2$ single crystal. We applied the temperature gradient $\bm{\nabla}T$ in a 45$^{\circ}$ rotated direction from the both principal crystallographic axes, as shown in Fig. 2(b).
Figure 7(a) shows the temperature dependence of the longitudinal temperature difference $\Delta T_{\mathrm{L}}$ using thermocouples connected at points 1 and 2, and the transverse temperature difference $\Delta T_{\mathrm{T}}$ using thermocouples connected at points 3 and 4.
In Fig. 7(b), we also show the temperature dependence of the potential difference measured along the longitudinal direction $\Delta V_{\mathrm{L}}=V_{12}$ and the transverse direction $\Delta V_{\mathrm{T}}=V_{34}$ where $V_{ij}$ is the potential difference between the thermocouple contact $i$ and $j$ in Fig. 2(b).
Although $|\Delta T_{\mathrm{T}}|$ is much smaller than $|\Delta T_{\mathrm{L}}|$ [Fig. 7(a)], $|\Delta V_{\mathrm{T}}|$ is larger than $|\Delta V_{\mathrm{L}}|$ [Fig. 7(b)], indicating that the majority of the longitudinal heat current is converted to the transverse electric field when the temperature gradient is applied to a direction between the $a$ and $c$ axes. Similar results are obtained in the sample~$\#$2, confirming the reproducibility of this experiment (see Fig. S8 in the Supplemental Material \cite{supplemental}).
Note that the temperature difference $|\Delta T_{\mathrm{L}}|$ and $|\Delta T_{\mathrm{T}}|$ were slightly enhanced below 30~K to evaluate small thermoelectric voltage at low temperatures.

When we assume a homogeneous temperature gradient,
the obtained thermoelectric voltage can be estimated as
\begin{align}
\Delta V_{\mathrm{L}}&=S_{xx}\Delta T_{\mathrm{L}}+\dfrac{l}{w}S_{xy}\Delta T_{\mathrm{T}},\\
\Delta V_{\mathrm{T}}&=S_{yy}\Delta T_{\mathrm{T}}+\dfrac{w}{l}S_{yx}\Delta T_{\mathrm{L}},
\end{align}
where $l$ and $w$ are the distance between the contacts of thermocouples along longitudinal and transverse direction respectively. Here, $S_{xx}$ and $S_{yy}$ are the diagonal components of the thermopower tensor in the $xy$ coordinate, where the $x$ and $y$ axes are parallel and perpendicular to the direction of $-\bm{\nabla} T$. In this work, we applied $-\bm{\nabla}T$ in an approximately 45$^{\circ}$ rotated direction from both the $a$ and $c$ axes, then $S_{xx}$ and $S_{yy}$ are equal to $\left(S_{aa}+S_{cc}\right)/2$. $S_{xy}$ and $S_{yx}$ are the off-diagonal terms of the thermopower tensor, which play a significant role in TTE conversion, and are equal to $\left(S_{aa}-S_{cc}\right)/2$. In materials with ADCP, $S_{xy}$ is enhanced because of the opposite polarity of the thermopower along different crystallographic axes, and then the large transverse voltage was obtained. The dotted curves in Fig. 7(b) show the calculated $\Delta V_{\mathrm{L}}$ and $\Delta V_{\mathrm{T}}$ using the right-hand side of Eqs. (3) and (4), respectively. 
The thermoelectric voltage from the wire leads is also included in the calculations.
These calculated values are in good agreement with the experimental results, although a slight deviation is seen below 50 K.

We also calculated the temperature and potential distribution using the finite element method to take into account the space-dependent gradient.
As explained in Appendix B, we use the experimental transport coefficient $S_{ii},\rho_{ii}$ and $\kappa_{ii}\,\left(i=a,c\right)$ at each temperature.
The calculated results of the temperature and the potential map at 300 K are shown in Figs. 7(c) and 7(d), respectively.
Clearly, the heat current $-\bm{\nabla}T$ along the $x$ axis is converted to the transverse potential difference along the $y$ axis. 
On the other hand, the contour plots at 30 K shown in Figs. 7(e) and 7(f) are significantly different from the plots at 300 K. The potential gradient is rather along the $c$ axis and inhomogeneous along the transverse direction due to the strong anisotropy in the electrical resistivity and thermal conductivity at low temperatures, as shown in Figs. 3(e) and 3(f). 
From the calculated temperature map, we obtain the transverse temperature difference $\Delta T_{\rm T}$ between two contact points along the transverse direction shown in Fig. 7(c). Likewise, from the potential map, we obtain the longitudinal potential difference $\Delta V_{\rm L}$ and the transverse one $\Delta V_{\rm T}$ between two contact points along the longitudinal and the transverse directions shown in Fig. 7(c), respectively.
The overall behavior of the calculated $\Delta T_{\mathrm{T}}$, $\Delta V_{\mathrm{T}}$, and $\Delta V_{\mathrm{L}}$ shown in Figs. 7(a) and 7(b) are close to the experimental results.

\begin{figure}[t]
\begin{center}
\includegraphics[width=9cm]{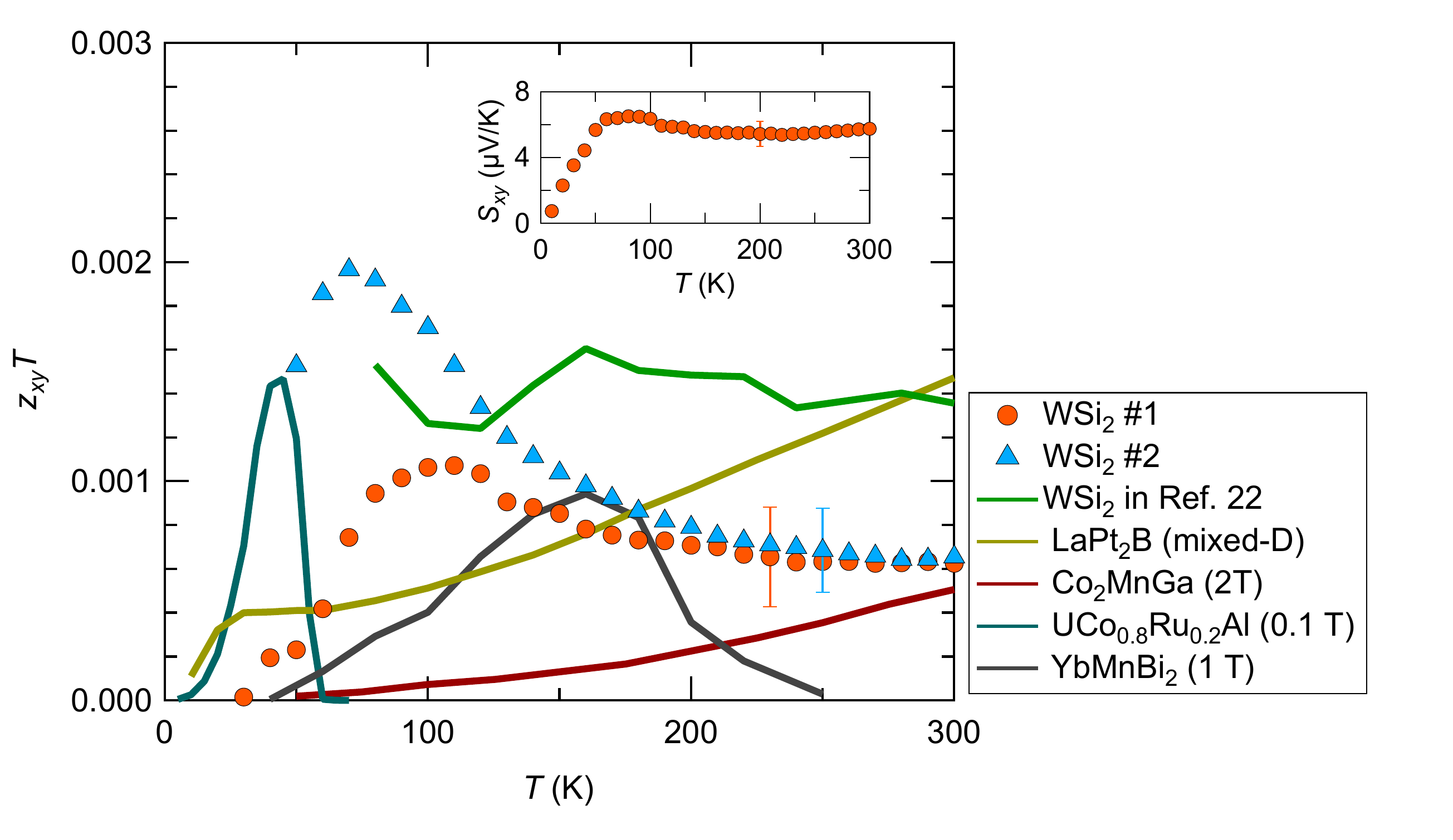}
\caption{The comparison of the temperature dependence of transverse dimensionless figure of merit $z_{xy}T$ of the mixed-dimensional (mixed-D) conductors and the ANE materials. WSi$_{2}$ of this work and previous study \cite{Koster2023} and LaPt$_{2}$B \cite{manako2024} are the mixed-dimensional conductors, while UCo$_{0.8}$Ru$_{0.2}$Al \cite{asaba2021colossal}, YbMnBi$_{2}$ \cite{pan2022giant} and Co$_{2}$MnGa \cite{sakai2018giant} are the typical examples of the ANE transverse thermoelectric materials. The inset shows the temperature dependence of $S_{xy}$ of sample $\#1$. The error bars represent the systematic error.}
\end{center}
\end{figure}

Figure 8 shows the temperature dependence of transverse dimensionless figure of merit $z_{xy}T$ of the mixed-dimensional conductors and the ANE materials. In our experimental conditions, $z_{xy}T$ is  given as \cite{tang2015p}
\begin{align}
    z_{xy}T=\dfrac{4S_{xy}^2T}{\left(\rho_{aa}+\rho_{cc}\right)\left(\kappa_{aa}+\kappa_{cc}\right)}.
\end{align}

The inset of Fig. 8 shows the temperature variation of $S_{xy}$, which is obtained as $S_{xy}=S_{yx}=\left(\Delta V_{\mathrm{T}}/w\right)/(\Delta T_{\mathrm{L}}/l)$, yielding 6 $\mathrm{\mu V/K}$ near room temperature. Moreover, $z_{xy}T$ of WSi$_{2}$ exhibits $\simeq1.0\times10^{-3}$ above 100 K, which is comparable to that of the ferromagnetic topological Heusler compound at room temperature \cite{sakai2018giant}. Note that the systematic error is evaluated using the error propagation from measured transport coefficients. Although the measured $z_{xy}T$ value has considerable uncertainty in general \cite{heremans2024thermoelectric}, this result indicates that the mixed-dimensional conductor such as WSi$_{2}$ and LaPt$_{2}$B exhibits a great TTE performance in a zero magnetic field and wide temperature range. Although the transverse thermopower of TTE materials is still an order of magnitude lower than the thermopower of conventional thermoelectric materials, such mixed-dimensional conductors are applicable to a heat flux sensor, which directly detects a heat flux perpendicular to the sensor plane. Monitoring the heat flux can provide important information for energy management, and the perpendicular heat flux sensor is actually demonstrated in the ANE materials \cite{tanaka2023roll}. WSi$_{2}$ single crystals are also expected to be used for similar applications without a magnetic field.

\section{Conclusions}
In this study, we evaluated the thermoelectric properties of WSi$_{2}$ single crystals below room temperature. The ADCP of the thermopower was confirmed and well understood by the inverse mass anisotropy of the mixed-dimensional Fermi surfaces. The sample-dependent transport properties were also figured out. Thermopower, electrical resistivity, and thermal conductivity exhibit a notable sample dependence, and the thermopower calculations taking the band-dependent scattering rate give a plausible interpretation for the sample-dependent transport properties of WSi$_2$ single crystals. The TTE conversion based on the ADCP was demonstrated by applying the temperature gradient to the direction rotated from the crystallographic axis in a WSi$_{2}$ single crystal. The FEM analysis reproduces the TTE effect in WSi$_{2}$. Our study indicates that the mixed-dimensional semimetal is a great candidate for the TTE materials.

\section*{Acknowledgements}
We appreciate R. Kurihara and H. Yaguchi for discussions and H. Manako for assistance.
This work was partly supported by JSPS KAKENHI Grant No. 22K20360, No. 22H01166, and No. 24K06945, and Research Foundation for the Electrotechnology of Chubu (REFEC, No. R-04102).

\section*{Appendix}

\subsection{Calculations of the transport properties}

To examine the ADCP and the sample-dependent thermopower, we also calculated the transport coefficients based on the linearized Boltzmann equations under the constant-relaxation time approximation.
From the obtained eigenvalues of the $n$-th band at the $\bm{k}$ point $\left(\varepsilon_{n,\bm{k}}\right)$, the transport distribution function tensor $L_{ij}\left(\varepsilon\right)$ is given as
\begin{align}
L_{ij}\left(\varepsilon\right)=\sum_{n}L_{ij,n}\left(\varepsilon\right)=\sum_{n}\sum_{\bm{k}}v_{i}v_{j}\tau_0\delta\left(\varepsilon-\varepsilon_{n,\bm{k}}\right),
\end{align}
where $L_{ij,n}\left(\varepsilon\right)$ is the partial transport distribution function tensor of $n$-th band, $v_{i}$ is the $i$-th component of the band velocity $\bm{v}=\dfrac{1}{\hbar}\bm{\nabla}_{k}\varepsilon_{n,\bm{k}}$, $\tau_0\left(=10^{-14}\,\mathrm{s}\right)$ is the relaxation time, and $\delta$ is the delta function.

Then, the transport coefficients are evaluated as a function of the chemical potential $\mu$ for the temperature $T$. The partial electrical conductivity tensor for $n$-th band $\sigma_{ij,n}\left(\mu\right)$ is
\begin{align}
\sigma_{ij,n}\left(\mu\right)=e^2\int_{-\infty}^{\infty}d\varepsilon\left(-\dfrac{\partial f_{0}}{\partial\varepsilon}\right)L_{ij,n}\left(\varepsilon\right)
\end{align}
where $e$ is the elementary charge and $f_{0}$ is the Fermi-Dirac distribution function. The total electrical conductivity tensor $\sigma_{ij}\left(\mu\right)$ is the sum of the $\sigma_{ij,n}\left(\mu\right)$.

Similarly, the partial Peltier conductivity tensor of $n$-th band $\left(\sigma S\right)_{ij,n}\left(\mu\right)$ is
\begin{align}
\left(\sigma S\right)_{ij,n}\left(\mu\right)=-\dfrac{e}{T}\int_{-\infty}^{\infty}d\varepsilon\left(-\dfrac{\partial f_{0}}{\partial\varepsilon}\right)\left(\varepsilon-\mu\right)L_{ij,n}\left(\varepsilon\right),
\end{align}
where $S_{ij}$ is the thermopower tensor. The total Peltier conductivity $\left(\sigma S\right)_{ij}\left(\mu\right)$ is the sum of the $\alpha_{ij,n}\left(\mu\right)$. The thermopower $S_{ii}$ for the $i$-axis direction is obtained as $S_{ii}=\left(\sigma S\right)_{ii}/\sigma_{ii}$.

The inverse mass tensor of $n$-th band is obtained using the calculated energy of the $n$-th band at $\bm{k}$ point $\varepsilon_{n,\bm{k}}$ as
\begin{align}
\left(\dfrac{1}{m^{\ast}}\right)_{ij}=\dfrac{1}{\hbar^2}\dfrac{\partial^{2}\varepsilon_{n,\bm{k}}}{\partial k_{i}\partial k_{j}}
\end{align}
indicating that the positive (negative) value of the inverse mass tensor expresses electron-like (hole-like) band curvature. The chemical potential dependence of the Hall coefficient $R_{\mathrm{H}}$ was calculated by using BoltzTraP2 code \cite{madsen2018boltztrap2,luo2023direction}.

\subsection{Field map}
Field map were calculated using the finite element method based on the transport equations \cite{Silk2008}. Transport equations are given as
\begin{align}
\bm{j}_{\mathrm{e}}&=\hat{\sigma}\left(-\bm{\nabla}\phi\right)+\hat{\sigma}\hat{S}\left(-\bm{\nabla}T\right)\label{eq:method_j_e},\\
\bm{j}_{\mathrm{Q}}&=\hat{\sigma}\hat{S}T\left(-\bm{\nabla}\phi\right)+\hat{\kappa}\left(-\bm{\nabla}T\right)\label{eq:method_j_Q},
\end{align}
where $\bm{j}_{\mathrm{e}}$ is the electrical current density, $\bm{j}_{\mathrm{Q}}$ is the thermal current density, $\hat{\sigma}$ is the electrical conductivity tensor, $\hat{S}$ is the thermopower tensor, and $\hat{\kappa}$ is the thermal conductivity tensor.
First, we calculated the temperature map based on Eq. \eqref{eq:method_j_Q}. 
In this calculation, we used the experimental values of $\hat{\sigma}$, $\hat{S}$, and $\hat{\kappa}$.
Assuming that the Peltier thermal current $\hat{\sigma}\hat{S}T\left(-\bm{\nabla}\phi\right)$ and Joule heat are negligible, we solved the following equation of continuity,
\begin{equation}
\bm{\nabla}\cdot\bm{j}_{\mathrm{Q}}=\bm{\nabla}\cdot\left[\hat{\kappa}\left(-\bm{\nabla}T\right)\right]=0.
\end{equation}
Here, we applied a constraint that the calculated longitudinal temperature difference $\Delta T_{\rm L}$ should equal to the experimental value. To realize the experimental setup shown in Fig. 2(b), the heat source and sink are located at the bottom-left and top-right corners in Figs. 7(c-f), respectively.
From the calculated temperature map, we obtain the transverse temperature difference $\Delta T_{\rm T}$ between two contact points along the transverse direction shown in Fig. 7(c). The obtained data of $\Delta T_{\rm T}$ are plotted as the open symbols in Fig. 7(a).

Next, we calculated the potential map based on the Eq. \eqref{eq:method_j_e}. The continuity equation is given as
\begin{equation}
\bm{\nabla}\cdot\bm{j}_{\mathrm{e}}=\bm{\nabla}\cdot\left[\hat{\sigma}\left(-\bm{\nabla}\phi\right)+\hat{\sigma}\hat{S}\left(-\bm{\nabla}T\right)\right]=0.
\end{equation}
From the calculated potential map, 
we obtain the longitudinal potential difference $\Delta V_{\rm L}$ between two contact points along the longitudinal direction shown in Fig. 7(c) and also obtain the transverse one $\Delta V_{\rm T}$ between two contact points along the transverse direction shown in Fig. 7(c). The obtained data of $\Delta V_{\rm L}$ and $\Delta V_{\rm T}$ are plotted as the open squares and open circles in Fig. 7(b), respectively.


%

\end{document}